\documentclass[final,5p,twocolumn]{elsarticle}

\usepackage[utf8x]{inputenc}
\usepackage{amssymb}
\usepackage{amsmath}
\usepackage{multirow}
\usepackage{longtable}

\usepackage{graphicx}
\usepackage{array}
\usepackage{dcolumn}
\usepackage{mathrsfs}
\usepackage{amsbsy}
\usepackage{units}
\usepackage{nicefrac}
\usepackage{slashed}

\newcommand{\Slash}[1]{\ooalign{\hfil/\hfil\crcr$#1$}}

\usepackage[usenames]{color}
\usepackage[colorlinks=true,linkcolor=blue,citecolor=blue,urlcolor=blue]{hyperref}

\definecolor{lred}{rgb}{1,0.90,0.7}

\journal{Physics Letters B}

\begin{document}

\begin{frontmatter}

\title{Pion cloud effects on baryon masses}

\author[gi]{H\`elios Sanchis-Alepuz\corref{HSA}}
\address[gi]{Institute of Theoretical Physics, Justus-Liebig University of Gie\ss en, \\Heinrich-Buff-Ring 16, 35392, Gie\ss en, Germany}
\author[gi]{Christian S. Fischer}
\author[gi]{Stanislav Kubrak}

\cortext[HSA]{helios.sanchis-alepuz@theo.physik.uni-giessen.de}

\begin{abstract}
In this work we explore the effect of pion cloud contributions
to the mass of the nucleon and the $\Delta$ baryon. To this end we solve
a coupled system of Dyson-Schwinger equations for the quark propagator,
a Bethe-Salpeter equation for the pion and a three-body Faddeev equation
for the baryons. In the quark-gluon interaction we explicitly resolve
the term responsible for the back-coupling of the pion onto the quark,
representing rainbow-ladder like pion cloud effects in bound states.
We study the dependence of the resulting baryon masses on the current
quark mass and discuss the internal structure of the baryons in terms of
a partial wave decomposition. We furthermore determine values for the 
nucleon and $\Delta$ sigma-terms. 
\end{abstract}


\end{frontmatter}

\section{Introduction}
\label{sec:intro}

The application of continuum functional methods to hadron physics phenomenology aims at the calculation of hadronic properties using the elementary degrees of freedom of Quantum Chromodynamics (QCD). In this framework mesons and baryons are considered as bound states of quarks and, hence, described by two-body Bethe-Salpeter equations (BSEs) and three-body Faddeev equations. These equations rely upon the knowledge of several QCD's Green's functions which are in turn solutions of Dyson-Schwinger equations (DSEs). The approach has the advantage that the origin of physical observables can be understood from the microscopic dynamics of quarks and gluons. Moreover, it is Poincar\'e covariant and is applicable at any momentum range.

As is well known, however, it is impossible to carry out this program exactly and truncations of both the DSEs and the bound state equations must be defined. The simplest one consistent with Poincar\'e covariance as well as constraints from chiral symmetry is the Rainbow-Ladder truncation (RL). Approximations of this kind have been extensively used in hadron calculations (see e.g. \cite{Eichmann:2013afa,Bashir:2012fs} for overviews) and turn out to be rather successful in reproducing, e.g., ground-state masses 
in selected channels.

There are, however, also severe limitations to the 
rainbow-ladder scheme. Consequently, much work has been 
invested in the past years on its extension towards more 
advanced approximations of the quark-gluon interaction. 
On the one hand, this may be accomplished directly by devising
improved \textit{ans\"atze} for the dressing functions of the 
quark-gluon vertex 
\cite{Fischer:2005en,Chang:2009zb,Chang:2010hb,heupel_new}.
On the other hand, it is promising to work with diagrammatic
approximations to the vertex DSE. While most studies so far concentrated 
on ($1/N_c$-subleading) Abelian contributions to the vertex (see e.g. 
\cite{Bender:1996bb,Watson:2004kd,Watson:2004jq,Bhagwat:2004hn,Matevosyan:2006bk}),
the impact of the $1/N_c$-leading, non-Abelian diagram on light meson
masses has been investigated in \cite{Fischer:2009jm}. In addition,
important unquenching effects in the quark-gluon interaction may 
be approximated by the inclusion of hadronic degrees 
of freedom \cite{Fischer:2007ze,Fischer:2008sp,Fischer:2008wy}. 
This is possible, since the vertex DSE can be decomposed on a diagrammatic
level into terms that are already present in the quenched theory and those 
involving explicit quark-loops. The latter ones can be expressed involving  
hadronic degrees of freedom. To leading order in the hadron masses, pion 
exchange between quarks is dominating these contributions. These pions are 
not elementary fields. Consequently, their wave functions need to be determined 
from their Bethe-Salpeter equation. 

Having explicit hadronic degrees of freedom in the system may also be very 
beneficial for phenomenological applications of the approach. Pion cloud effects 
are expected to play an important role in the low momentum behavior of 
form factors and hadronic decay processes of baryons
\cite{Thomas:1981vc,Miller:2002ig,Ramalho:2008dp,Cloet:2012cy,Eichmann:2011vu,%
Eichmann:2011aa,Sanchis-Alepuz:2013iia,NDg}. Within the covariant BSE-approach, 
the influence of pion back-coupling effects in the mass and decay 
constants of the pion itself and other light mesons has been studied in \cite{Fischer:2008wy}. 
In the present work, we take this framework one step further and extend it 
to the covariant three-body calculations of nucleon and delta masses 
\cite{Eichmann:2009qa,Eichmann:2009en,SanchisAlepuz:2011jn}. 

This letter is organized as follows: in Section \ref{sec:framework} we review the main elements of the DSE/BSE framework and define the truncations and model used in this work. We present and discuss the results of our calculations in Section \ref{sec:results}. Finally, some concluding remarks are made in Section \ref{sec:summary}.

\section{Covariant three-body equation}
\label{sec:framework}

\begin{figure*}[hbtp]
 \begin{center}
  \includegraphics[width=0.86\textwidth,clip]{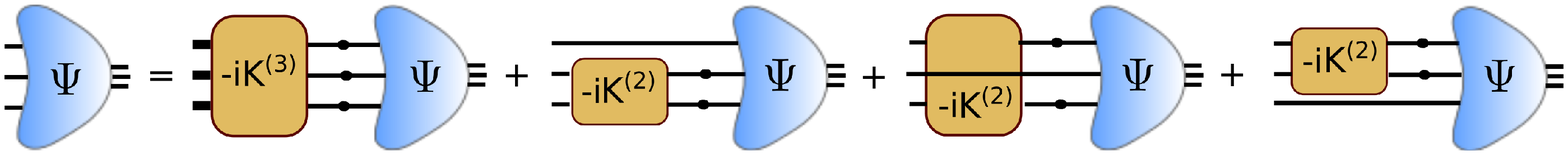}
 \end{center}
 \caption{Diagrammatic representation of the three-body Bethe-Salpeter equation.}\label{fig:faddeev_eq}
 \begin{center}
  \includegraphics[width=0.53\textwidth,clip]{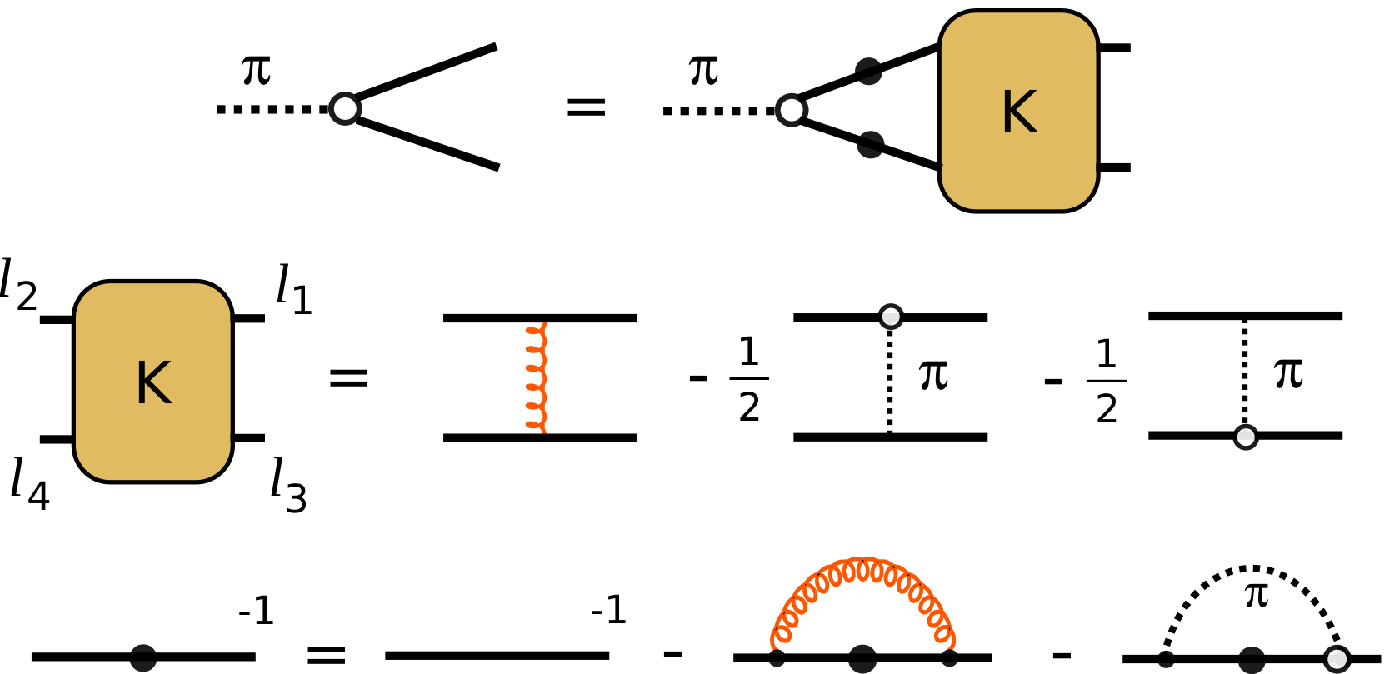}
 \end{center}
 \caption{Diagrammatic representation of the pion BSE, two-body kernel and quark DSE when pion-exchange contributions are included. Big and small blobs represent full and bare vertices, respectively.}\label{fig:system_with_pion}
\end{figure*}

The mass and internal structure of baryons are given, in a covariant
Faddeev approach, by the solutions of the three-body equation
\begin{equation}\label{eq:3bBSEcompact}
\Psi = -i\widetilde{K}^{(3)}~G_0^{(3)}~\Psi + \sum_{a=1}^3 -i\widetilde{K}_{(a)}^{(2)}~G_0^{(3)}~\Psi\,,
\end{equation}
where $\widetilde{K}^{(3)}$ and $\widetilde{K}^{(2)}$ are the three- and two-body interaction kernels, respectively, and $G_0$ represents the product of three fully-dressed quark propagators $S$. We used here a compact notation where indices have been omitted and we assume that discrete and continuous variables are summed or integrated over, respectively. 
The spin-momentum part of the full amplitude $\Psi$ depends on 
the total and two relative momenta of the three valence quarks inside 
the baryon. As discussed in more detail in Section \ref{subsec:internal_composition}, 
this amplitude contains all possible spin and orbital angular momentum contributions.

The quark propagators are obtained from their respective DSE
\begin{equation}\label{eq:quarkDSE}
 S^{-1}(p)=S^{-1}_0(p)-Z_{1f}\int_q \Gamma^\nu_{gqq,0}
D_{\mu\nu}(p-q)\Gamma^\nu_{gqq}(p,q)S(q)\,\,,
\end{equation}
where the integration over the four-momentum $q$ is abbreviated by
$\int_q \equiv \int d^4 q/ {(2\pi)^4}$, $S_0$ is the (renormalized) 
bare propagator with its inverse given by
\begin{equation}\label{eq:bare_prop}
 S^{-1}_0(p)=Z_2\left(i\Slash{p}+m_q\right)\,\,,
\end{equation}
with bare quark mass $m_q$, whereas  
\begin{equation}\label{eq:prop}
 S^{-1}(p)=i\Slash{p} A(p^2) + B(p^2)\,\,,
\end{equation}
denotes the inverse dressed propagator. The renormalisation point invariant 
running quark mass $M(p^2)$ is defined by the ratio of the scalar 
quark dressing function $B(p^2)$ and the vector dressing function 
$A(p^2)$: $M(p^2) = B(p^2)/A(p^2)$.
$\Gamma^\nu_{gqq}$ is the full quark-gluon vertex with 
its bare counterpart $\Gamma^\nu_{gqq,0}$, $D^{\mu\nu}$ is the full
gluon propagator and $Z_{1f}$ and $Z_2$ are renormalization constants.

To solve the system formed by equations (\ref{eq:3bBSEcompact}) and (\ref{eq:quarkDSE}) one needs to know the interaction kernels and the full quark-gluon vertex. The latter could in principle be obtained from the infinite system of coupled DSEs of QCD. In practice, however, this system has to be truncated into something manageable, which implies that educated \textit{ans\"atze} have to be used for the Green's functions one is not solving for. The interaction kernels, in contrast, do not appear directly in the system of QCD's DSEs. 
In the quark-antiquark channel, a connection of those with the quark-gluon interaction is established
via the axial-vector Ward-Takahashi identity, which ensures the correct implementation of chiral 
symmetry in the bound state equations \cite{Munczek:1994zz,Maris:1997hd}. In turn, it is natural from
a systematic point of view to treat the interaction kernels in the quark-quark channels on a similar
approximation level, such that both kernels are fixed once the approximation of the
quark-gluon interaction is specified. This will be detailed below.

\subsection{Rainbow-Ladder truncation}
\label{subsec:RL}

The simplest and most commonly used ansatz for the quark-gluon and quark-quark interactions is the Rainbow-Ladder (RL) truncation. Here, only the tree-level flavor, color and Lorentz structures are kept for the quark-gluon vertex, so that the quark DSE reads
\begin{equation}\label{eq:quarkDSERL}
 S^{-1}_{\alpha\beta}(p)=S^{-1}_{0,\alpha\beta}(p)-\int_q
\widetilde{K}^{RL}_{\alpha\alpha'\beta'\beta}(k)S_{\alpha'\beta'}(q)\,\,,
\end{equation}
with momentum $k=p-q$ and kernel
\begin{equation}\label{eq:RLkernel}
	\widetilde{K}^{RL}_{\alpha\alpha'\beta'\beta}(k)= -4\pi C~Z_2^2
~\frac{\alpha_{\textrm{eff}}(k^2)}{k^2}~
	T_{\mu\nu}(k)~\gamma^\mu_{\alpha\alpha'}  \gamma^\nu_{\beta'\beta}\,\,.
\end{equation}
Here $Z_2$ denotes the quark renormalization constant, $T_{\mu\nu}(k)$ the transverse projector
\begin{equation}\label{eq:def_transverse_proj}
 T_{\mu\nu}(k)=\delta_{\mu\nu}-\frac{k_\mu k_\nu}{k^2}\,\,,
\end{equation}
and $C=4/3$ the resulting color factor for quarks in fundamental representation. 
The effective coupling $\alpha_{\textrm{eff}}$ 
combines the non-perturbative dressing of the gluon propagator and the $\gamma_\mu$-structure of the vertex.
At large momenta, it is constrained by perturbation theory, whereas at low momenta we have to supply a model. 
In this work we use the model proposed in \cite{Maris:1997tm,Maris:1999nt}
\begin{flalign}\label{eq:MTmodel}
\alpha_{\textrm{eff}}(q^2) {}=&
 \pi\eta^7\left(\frac{q^2}{\Lambda^2}\right)^2
e^{-\eta^2\frac{q^2}{\Lambda^2}}\nonumber\\ &+{}\frac{2\pi\gamma_m
\big(1-e^{-q^2/\Lambda_{t}^2}\big)}{\textnormal{ln}[e^2-1+(1+q^2/\Lambda_{QCD}
^2)^2]}\,, 
\end{flalign}
where for the anomalous dimension we use $\gamma_m=12/(11N_C-2N_f)=12/25$,
corresponding to $N_f=4$ flavors and $N_c=3$ colors. We fix the QCD scale to
$\Lambda_{QCD}=0.234$ GeV and the scale $\Lambda_t=1$~GeV is introduced for technical reasons and has no impact on
the results. The interaction strength is characterized by an energy
scale $\Lambda$ and the dimensionless parameter $\eta$ controls the width of the interaction. 
They have to be fixed by experimental input, see Section \ref{sec:results}. 

The quark-antiquark kernel in the pion Bethe-Salpeter equation (BSE) has to match the interaction
model in the quark-DSE such as to guarantee the Goldstone-boson property of the pion
in the chiral limit. This is encoded in the axialvector Ward-Takahashi identity (axWTI).
In the rainbow-ladder truncation, the quark-antiquark kernel in the BSE is then also given
by Eq.~(\ref{eq:RLkernel}). The corresponding kernel describing the interaction between two
quarks can be obtained via crossing symmetry. For our rainbow-ladder scheme this results in 
the same expression Eq.~(\ref{eq:RLkernel}) with modified color factor $C=-2/3$. For diquarks,
such a kernel together with its extensions has been explored e.g. in \cite{Bender:1996bb}, whereas
in the context of the three-body Faddeev equations first results have been reported 
in \cite{Eichmann:2009qa,Eichmann:2009en,SanchisAlepuz:2011jn}. In the latter studies, the 
three-body irreducible interactions between the three quarks have been neglected. We adopt the
same framework in this work. 

The three-body Faddeev-equation then reduces to
%
\begin{flalign}\label{eq:faddeev_eq}
&\hspace*{-5mm}\Psi_{\alpha\beta\gamma\mathcal{I}}(p,q,P) =~~~~~~~~~~~~~~~~~~~~~~~~~~~~~~~~~~~~~~~~~~~\nonumber\\
  \int_k & \left[
\widetilde{K}^{RL}_{\beta\beta'\gamma\gamma'}(k)~S_{\beta'\beta''}(k_2)
S_{\gamma'\gamma''}(\tilde{k}_3)~
\Psi_{\alpha\beta''\gamma''\mathcal{I}}(1,P)\right.\nonumber\\
& + \left.
\widetilde{K}^{RL}_{\alpha\alpha'\gamma\gamma'}(-k)~S_{\gamma'\gamma''}(k_3)
S_{\alpha'\alpha''} (\tilde{k}_1)~
\Psi_{\alpha''\beta\gamma''\mathcal{I}}(2,P)
\right. \nonumber\\
& + \left. 
\widetilde{K}^{RL}_{\alpha\alpha'\beta\beta'}(k)~S_{\alpha'\alpha''}(k_1)
S_{\beta'\beta''}(\tilde{k}_2)~
\Psi_{\alpha''\beta''\gamma\mathcal{I}}(3,P)\right]\,\,,
\end{flalign}
where the generic index $\mathcal{I}$ in $\Psi$ refers to the bound state and the first
three Greek indices refer to the valence quarks \cite{Eichmann:2009qa,Eichmann:2009en,SanchisAlepuz:2011jn}. 
The Faddeev amplitudes depend on the total baryon momentum $P$ and two
relative momenta $p$ and $q$
\begin{align}\label{eq:defpq}
        p &= (1-\zeta)\,p_3 - \zeta (p_1+p_2)\,, &  p_1 &=  -q -\dfrac{p}{2} +
\dfrac{1-\zeta}{2} P\,, \nonumber\\
        q &= \dfrac{p_2-p_1}{2}\,,         &  p_2 &=   q -\dfrac{p}{2} +
\dfrac{1-\zeta}{2} P\,, \nonumber\\
        P &= p_1+p_2+p_3\,,                &  p_3 &=   p + \zeta  P\,\,,\nonumber\\
\end{align}
with $p_1$, $p_2$ and $p_3$ the quark momenta and $\zeta$ a free momentum partitioning parameter, which is chosen to be $\zeta=1/3$ for numerical convenience. The quark propagators depend on 
the internal quark momenta $k_i=p_i-k$ and $\tilde{k}_i=p_i+k$, with $k$ the gluon momentum. Similarly, the internal relative momenta 
$(j,P) \equiv (p^{(j)},q^{(j)},P)$
for each of the three terms in the Faddeev equation are
\begin{align}\label{internal-relative-momenta}
p^{(1)} &= p+k,& p^{(2)} &= p-k,& p^{(3)} &= p,\nonumber\\
q^{(1)} &= q-k/2,& q^{(2)} &= q-k/2, & q^{(3)} &= q+k\,\,.\nonumber\\
\end{align}

\subsection{Beyond Rainbow-Ladder: Pion exchange}
\label{subsec:pion_exchange}

As discussed above, in this work we follow the framework of 
Refs.~\cite{Fischer:2007ze,Fischer:2008sp,Fischer:2008wy}. We will briefly summarize
the approximation scheme here, referring the reader to the original work for 
all technical details. There, the Dyson-Schwinger
equation for the quark-gluon vertex has been analyzed in detail and terms
representing (off-shell) hadronic contributions to the full vertex have been 
identified. To leading order in a $1/N_c$-expansion and in the mass of the 
exchanged hadrons, the dominant effect is that of the exchange of one pion.
Once this contribution to the vertex is inserted into the quark-DSE, the 
resulting two-loop diagram has been approximated by the one-loop graph shown
on the right hand side of the quark-DSE in Fig.~\ref{fig:system_with_pion}:
besides the well-known rainbow-ladder gluonic part of the quark-DSE a 
diagram representing the emission and subsequent absorption of a pion has 
appeared. Here, the coupling of the pion to the quark is given by a bare
pseudoscalar vertex and a full pion Bethe-Salpeter amplitude. Note, however, that
in general also the choice of two dressed vertices is possible and it is not 
clear a priori, which of the two choices is the better approximation of the
original two-loop diagram. In \cite{Fischer:2008wy} the choice
with one bare vertex led to satisfactory results in the vector-meson
sector and we will therefore adopt this also here.

The quark-DSE then reads
\begin{eqnarray}
 S^{-1}_{\alpha\beta}(p)=S^{-1}_{0,\alpha\beta}(p)&-&\int_q
\widetilde{K}_{\alpha\alpha'\beta'\beta}^{RL}(k)S_{\alpha'\beta'}(q)\nonumber\\ &-&\int_q
\widetilde{K}_{\alpha\alpha'\beta'\beta}^{\textrm{pion}}(k)S_{\alpha'\beta'}(q)\,\,,
\label{eq:quarkDSE_pi}
\end{eqnarray}
with the rainbow-ladder kernel $\widetilde{K}^{RL}$ from the previous section and with the 
additional kernel for the quark-pion interaction given by
\begin{align}\label{eq:PiKernel} 
  \widetilde{K}_{\alpha\alpha'\beta\beta'}^{\textrm{pion}}&(l_1,l_2,l_3,l_4;P)={} \nonumber\\
  &\frac{1}{2}
      [\Gamma^j_{\pi}]_{\alpha\alpha'}\left(\frac{l_1+l_2}{2};P\right)
      [Z_2 \tau^j \gamma_5]_{\beta\beta'}
       D_{\pi}(P) \nonumber\\
 & +\frac{1}{2}
      [Z_2 \tau^j \gamma_5]_{\alpha\alpha'}
      [\Gamma^j_{\pi}]_{\beta\beta'}\left( \frac{l_3+l_4}{2};P\right)
      D_{\pi}(P)    
      \,.
\end{align}
Here $\Gamma^j_{\pi}(p,P)$ is the pion Bethe-Salpeter amplitude, with relative momentum $p$,
total momentum $P$ and $l_{1..4}$ are the incoming and outgoing quark momenta as specified 
in the second line of Fig.\ref{fig:system_with_pion}. 
The pion propagator is given by
\begin{equation}\label{eq:pion_propagator}
 D_{\pi}(P)=\frac{1}{M_{\pi}^2+P^2}
\end{equation}
and $Z_2 \tau^j \gamma_5$ is the bare pion-quark vertex. 
The pion Bethe-Salpeter amplitude as obtained from its BSE is given by
\begin{eqnarray}
\Gamma_\pi^j(p;P) &=& \tau^j \gamma_{5} \left[E_\pi(p;P)- i \slashed P F_\pi(p;P) \right.\nonumber\\
&&\left.
- i \slashed p G_\pi(p;P) - \left[\slashed P ,\slashed p \right] H_\pi(p;P)\right]\,.\label{eq:pseu}
\end{eqnarray}
In principle, the back-coupling of the pion onto the quark is also
governed by this amplitude and thus one needs to solve the coupled
system of the quark-DSE and pion-BSE. In order to simplify this 
tremendous numerical task we only employ the leading amplitude $E_\pi(p;P)$
for the internal pion and neglect contributions from $F_\pi, G_\pi$ and
$H_\pi$. From a comparison of the relative size of these amplitudes we
estimate a total error of less than five percent due to this 
approximation.\footnote{The reasoning is as follows: It has been shown
already in Ref.~\cite{Maris:1997tm} that this approximation results in an error
of about twenty percent in the resulting pion mass. We take this as measure 
for the expected corresponding error in the pion exchange kernel. Since the 
inclusion of the pion back-reaction results in mass shifts of the baryons 
of at most twenty percent, we consequently expect a twenty percent error on 
this twenty percent which results in a total error less than five percent.}
For the external pion amplitude in its BSE we determine all four tensor 
components in Eq.~(\ref{eq:pseu}) without further approximations. 

The pion-exchange part of the interaction kernel represents an explicit unquenching
effect in the quark-gluon vertex. It is clear that the problem of determining 
the remaining parts of the vertex as well as the fully dressed gluon propagator
remains. This is an ongoing effort with much progress in recent years. On the
level of the Faddeev equation, however, the numerical effort involved in
beyond-rainbow-ladder calculations is extremely large, so that so far no such
study is available. We therefore also resort to a rainbow-ladder kernel representing
the remaining parts of the interaction. 

The resulting three-body equation (\ref{eq:faddeev_eq}) is formally still of ladder
type when the pion exchange is included and explicitly given by
\begin{flalign}\label{eq:faddeev_eq_pi}
&\hspace*{-5mm}\Psi_{\alpha\beta\gamma\mathcal{I}}(p,q,P) =~~~~~~~~~~~~~~~~~~~~~~~~~~~~~~~~~~~~~~~~~~~\nonumber\\
  \int_k & \left[
\widetilde{K}_{\beta\beta'\gamma\gamma'}(k)~S_{\beta'\beta''}(k_2)
S_{\gamma'\gamma''}(\tilde{k}_3)~
\Psi_{\alpha\beta''\gamma''\mathcal{I}}(1,P)\right.\nonumber\\
& + \left.
\widetilde{K}_{\alpha\alpha'\gamma\gamma'}(-k)~S_{\gamma'\gamma''}(k_3)
S_{\alpha'\alpha''} (\tilde{k}_1)~
\Psi_{\alpha''\beta\gamma''\mathcal{I}}(2,P)
\right. \nonumber\\
& + \left. 
\widetilde{K}_{\alpha\alpha'\beta\beta'}(k)~S_{\alpha'\alpha''}(k_1)
S_{\beta'\beta''}(\tilde{k}_2)~
\Psi_{\alpha''\beta''\gamma\mathcal{I}}(3,P)\right]\,\,,
\end{flalign}
with $\widetilde{K} = \widetilde{K}^{RL}-\widetilde{K}^{pion}$. The elements 
needed for the equation are the solutions of the system of equations depicted 
in Fig.~\ref{fig:system_with_pion}. In order to numerically solve the coupled 
system of equations for the quark DSE and pion BSE as well as the Faddeev equation 
we use standard numerical methods. In particular we employ a Cauchy-contour method
to solve the DSE and BSEs for the complex momenta needed in the Faddeev-equation.

\begin{figure*}[t]
 \begin{center}
  \includegraphics[width=0.45\textwidth,clip]{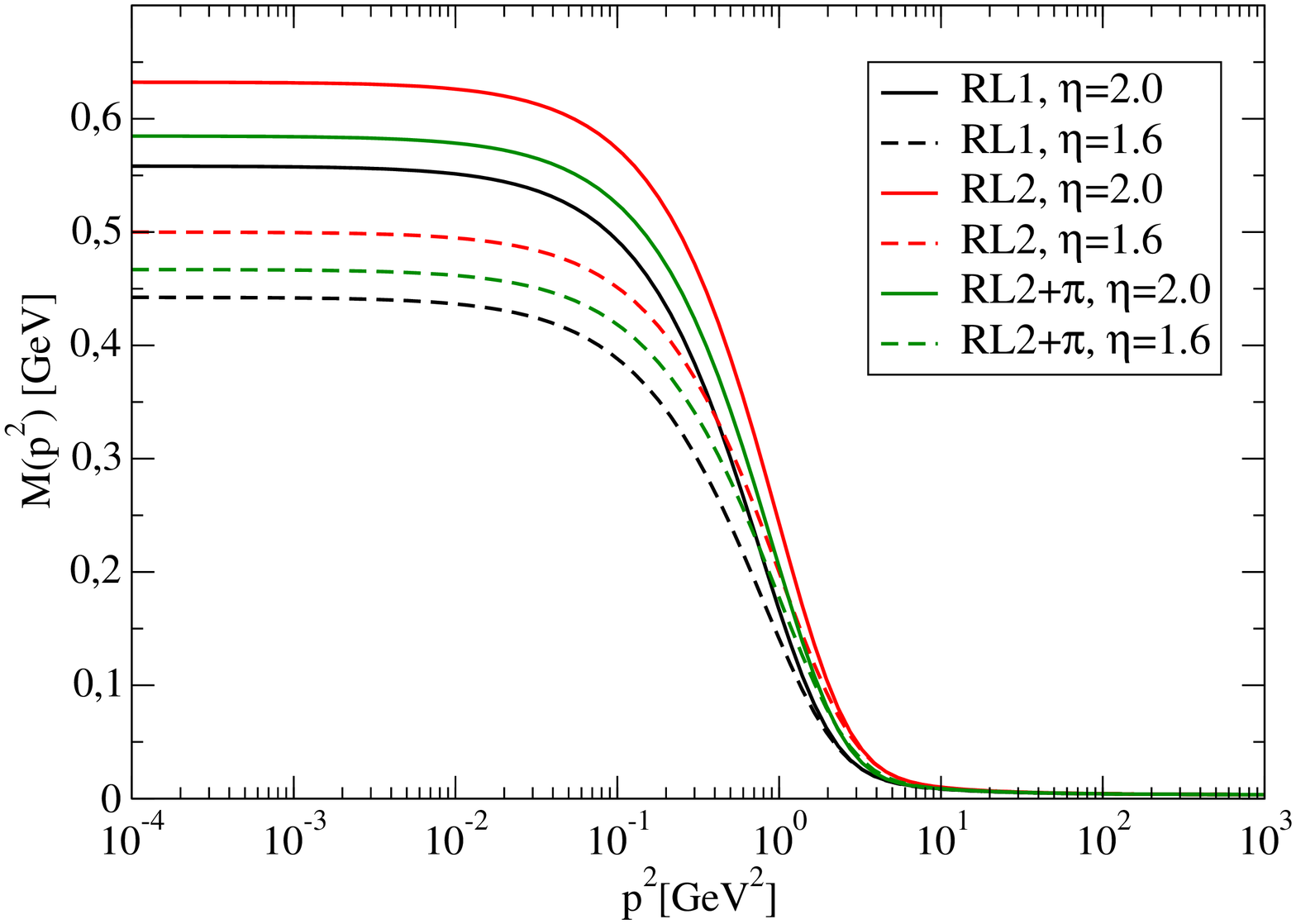}
 \end{center}
 \caption{(color online). Quark mass function as function of the squared momentum.}\label{fig:GMOR}
\end{figure*}

A couple of comments on the pion-exchange kernel are in order.
First, we wish to emphasize that the complete interaction kernel consisting of
the rainbow-ladder gluonic diagram and the pion exchange diagram does
satisfy the axial-vector Ward-Takahashi identity. This can be demonstrated 
analytically \cite{Fischer:2007ze,Fischer:2008wy} and holds even with the
approximation of the exchanged pion's Bethe-Salpeter amplitude introduced above. As a result,
using this interaction kernel one obtains a pseudoscalar Goldstone boson
in the chiral limit (from the pion Bethe-Salpeter equation) and the 
Gell-Mann-Oakes-Renner relation holds at the
physical point \cite{Fischer:2007ze,Fischer:2008wy}. 

Second, note that the pion exchange contribution originally arises as an
approximation of hadronic contributions to the quark-antiquark interaction
in the quark-gluon vertex \cite{Fischer:2007ze}. These contributions are
off-shell. Consequently, their large momentum behavior is not correctly
represented by the on-shell-approximation used in our pion kernel. Various
\textit{ans\"atze} for the off-shell continuation have been explored in the literature, 
see e.g. the discussion in Ref.~\cite{Nyffeler:2009tw} in the context of 
hadronic contributions to the anomalous magnetic moment of the muon. While 
these \textit{ans\"atze} may give some guidance, they are not unique and a satisfactory
solution of this problem involves solving the full four-body T-matrix. This
is left for future work. Since in this work we are mainly interested in 
the low-momentum behavior of the kernel, where the on-shell approximation is 
expected to be good, we remain within the on-shell approach framework of 
Refs.~\cite{Fischer:2007ze,Fischer:2008sp,Fischer:2008wy}.

Finally, note that the interaction kernel Eq.~(\ref{eq:PiKernel}) is not the 
full story in terms of diagrams. If the kernel were derived by the usual 
'cutting of diagrams' procedure as e.g. in a 2PI approach \cite{Munczek:1994zz}, 
a diagram would appear containing two internal pions. Such a diagram 
contains the important physics of opening up two-pion decay channels for certain
kinematics, relevant for example in the vector-meson sector. At present the resulting 
two-loop diagrams in the quark-antiquark interaction have not been addressed in the
DSE/BSE approach due to the numerical complexity involved. Instead, in 
\cite{Fischer:2007ze,Fischer:2008wy} the simpler one-pion exchange kernel has
been devised together with an appropriate choice of momentum arguments in the 
internal pion's Bethe-Salpeter amplitudes such that the ladder exchange 
kernel satisfies the axWTI. While a more complete approach finally has 
to deal with the two-loop diagram, in this exploratory calculation we will 
resort to the ladder contribution only.

\section{Results and Discussion}
\label{sec:results}

\begin{table*}[t]
 \begin{center}
 \small
\renewcommand{\arraystretch}{1.2}
  \begin{tabular}[h]{|c||c|c|c|c|}\hline
  [GeV]     &  RL1     & RL2      & RL2 + $\pi$ & Exp. \\ \hline\hline
 $m_\pi$  	& 0.138 (1)& 0.144 (1)& 0.138 (1)   & 0.140\\ \hline
 $f_\pi$  	& 0.093 (1)& 0.098 (1)& 0.093 (1)   & 0.093\\ \hline
 $\langle q\bar{q}\rangle^{1/3}_{\mu=19 \,\text{GeV}}$
          	& 0.281 (2)& 0.300 (3)& 0.280 (3)    &      \\ \hline\hline
 $m_N$      & 0.94 (1) & 1.01 (3) & 0.86 (1)    & 0.94 \\ \hline
 $m_\Delta$ & 1.23 (1) & 1.36 (1) & 1.30 (3)    & 1.23 \\ \hline
\end{tabular}
\caption{Nucleon and Delta masses as well as pion mass, decay constant and the chiral condensate 
using the rainbow-ladder truncation only (RL1),
rainbow-ladder with the refitted effective interaction (RL2) and including the pion 
cloud corrections corrections (RL2 + $\pi$). 
We give the central value of the bands corresponding to a variation of 
$\eta$ between $1.6 \le \eta \le 2.0$ with the halfwidth of the bands added in brackets. 
We compare also with experimental values.\label{tab:masses}}
 \end{center}
\end{table*}

\begin{figure*}[t]
\begin{center}
  \includegraphics[width=0.82\textwidth,clip]{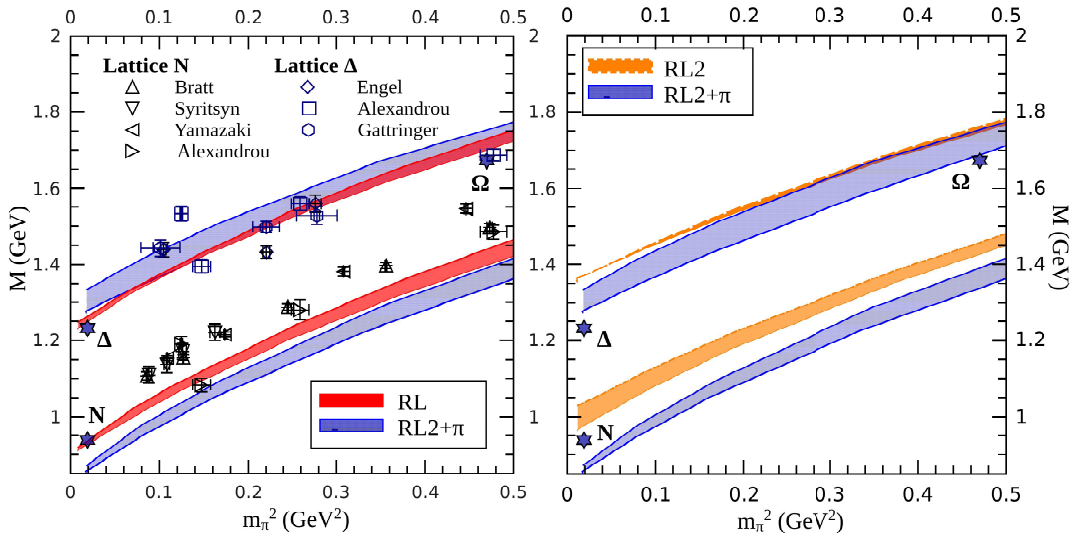}
 \end{center}
 \caption{(color online). Evolution of the nucleon and delta mass with respect to the pion mass squared. \textit{Left panel}: We plot the results for pure RL1 and for RL2 with pion exchange. We also compare with a selection of (unquenched) 
lattice data \cite{Alexandrou:2006ru}-\cite{Gattringer:2008vj}. \textit{Right panel}: We compare the results for RL2 only and RL2 with pion exchange. Stars denote the physical nucleon and delta mass.
The shaded bands correspond to a variation of the interaction parameter $\eta$ between
$1.6 \le \eta \le 2.0$, with $\eta=1.6$ corresponding to the upper limit of the bands.}\label{fig:mass_panel}
\end{figure*}

To proceed with the calculations we must fix the two parameters $\Lambda$ 
and $\eta$ of the interaction (\ref{eq:MTmodel}) as well as the 
current-quark masses. This is conveniently done by using the 
experimental values for the pion decay constant $f_\pi$ and the pion mass $m_\pi$
as benchmark. The pion decay constant is largely insensitive to the 
current quark mass, which is consequently fixed by the physical pion mass.
On the other hand, the parameter $\Lambda$ corresponds to an interaction 
scale, and is therefore in one-to-one relation with $f_\pi$. Furthermore,
it has been noted that the pion decay constant can only be reproduced by 
a range of values of $\eta$ between $1.6$ and $2.0$ (see, e.g. 
\cite{Eichmann:2011vu,Krassnigg:2009zh}). 
For the pure RL interaction $\widetilde{K}^{RL}$ the resulting values for $\Lambda$ and the quark mass
are $\Lambda=0.72$~GeV and $m_{u/d}(\mu^2)=3.7$~MeV; we denote this case by RL1.
Since the pion back-reaction is not taken into account explicitly in this case,
its effects are, to some extent, encoded implicitly in the parameters (in particular the scale) 
of the interaction. This is different for the pion corrected kernel 
$\widetilde{K} = \widetilde{K}^{RL}-\widetilde{K}^{pion}$. Since pion cloud effects are 
now treated explicitly, $\widetilde{K}^{RL}$ describes 
the interactions in the bound state's quark-core only. As a result, the 
interaction range of this part of the kernel (in coordinate space) is expected to decrease, which in turn means that 
$\Lambda$ should increase \cite{Thomas:1981vc}. This is indeed what we observe: for the 
pion-corrected kernel we need $\Lambda=0.84$~GeV to reproduce $f_\pi$ with 
$\eta \in [1.6,2.0]$. The quark mass $m_{u/d}(\mu^2)=3.7$~MeV remains the same.
We use the label RL2 for the RL part of this truncation. 
The renormalisation scale in all cases is chosen to be $\mu^2= (19 \,\mbox{GeV})^2$.

The resulting quark mass functions are displayed in Fig.~\ref{fig:GMOR}. For the 
two setups fixed by physical input, RL1 and RL2+$\pi$,
we find very similar mass functions with a difference in $M(0)$ of less than 
five percent. The quark-core setup $RL2$ generates slightly larger quark masses.
In general, the quark mass function encodes dynamical chiral symmetry breaking and 
nicely displays the transition from the low momentum notion of a constituent quark 
mass to the high momentum notion of a running current quark mass. Although the 
quark mass function is a renormalisation group invariant it is not, however, a 
gauge invariant quantity and therefore not directly observable. The chiral properties 
of our framework are also encoded in the dependence of the pion mass from the 
current quark mass. We explicitly checked the Gell-Mann-Oakes-Renner relation
for all setups and find that it holds within the numerical accuracy of 2 \%, 
as expected from the axWTI. The corresponding numbers are given in 
Tab.~\ref{tab:masses}.

\subsection{Nucleon and Delta masses and Sigma terms}

The calculated masses of the Nucleon and the Delta, with and without the 
pion-exchange kernel, are shown in Tab.~\ref{tab:masses}. In the RL1 framework
one observes very good agreement with the experimental 
mass values. However, as shown in Ref.~\cite{Eichmann:2011vu,Eichmann:2011pv}, the 
internal structure of the nucleon as probed by electromagnetic as well as axial 
and pseudoscalar currents is not well represented at low momenta due
to missing explicit pion cloud effects. These are included (within the limits of 
our truncation) in the RL2 + $\pi$-calculation. For comparison we also display 
results for the purely gluonic rainbow-ladder part of this truncation (RL2), which
represents a quark-core calculation of the nucleon mass with stripped pion cloud.
As a result we find substantial pion cloud effects in the nucleon. Compared with 
the quark-core part (RL2) the nucleon mass is reduced by about $150$ MeV in the 
full calculation (RL2+$\pi$). Comparing RL2+$\pi$ with RL1, which both reproduce
the physical pion mass and decay constant we still find pion cloud effects of the
order of $80$ MeV. This sizable mass shift for the nucleon at the physical
point agrees qualitatively with other estimates in the literature, see e.g. 
\cite{Young:2002cj} and references therein. The corresponding mass shift in the
$\Delta$-isobar is much smaller and behaves differently. Comparing RL2 and RL2+$\pi$
we find a decrease of the $\Delta$-mass by about $60$ MeV, which is less than half
the size of the corresponding shift in the nucleon. However, when comparing with
RL1, we even find an increase in the $\Delta$-mass by about $70$ MeV. This is a
result of the different interaction scale $\Lambda$ in the two setups, which
was necessary to reproduce the physical pion decay constant correctly. As a result
we find a mass shift of different sign for the $\Delta$ than for the nucleon.

The evolution of the baryon masses as a function of $m_\pi^2$ (or, equivalently, 
with respect to the current-quark mass), is displayed in Fig.~\ref{fig:mass_panel},
where we also display corresponding lattice data \cite{Alexandrou:2006ru}-\cite{Gattringer:2008vj}.
In general, we observe that the inclusion of pion cloud effects increases the mass 
splitting between the nucleon and the $\Delta$ considerably. Although the size of 
this increase may be too large, its qualitative behavior is in agreement with
well-known results in the literature \cite{Thomas:1981vc}. Including the pion
cloud effects, the excellent agreement of the pure rainbow-ladder calculation RL1
with experiment is spoiled and we are left with discrepancies for the nucleon
and the $\Delta$ on the ten percent level. Whereas the mass evolution for the 
$\Delta$ is not too far away from the corresponding lattice results, the one
for the nucleon is shifted by 10-20 percent for all pion masses, although the slope 
of the evolution is more or less correct.\footnote{When comparing our results with 
quenched and unquenched lattice data, 
one needs to be aware of different procedures of fixing the scales. Whereas in 
Ref.~\cite{Alexandrou:2009hs} both scales are matched to the physical point, in
Ref.~\cite{Young:2002cj} great care has been taken to exclude chiral contaminations
from the scale fixing procedure by matching to (short-range) static-quark forces.
The latter procedure resembles our comparison between RL2 and RL2 + $\pi$. 
However, since our quark-core calculation (RL2) is not in one-to-one 
relation with a quenched result (due to missing $\eta$-hairpin contributions)
we refrain from a detailed comparison with quenched data.}
In general, however, the quantitative discrepancies of our approach with the lattice 
results indicate missing structure such as gluon self interaction effects in the
two-body kernels (see \cite{Fischer:2009jm} for a study of these in the meson 
sector), genuine three-body interactions (also mediated by gluon self interaction
contributions) and potential deficiencies in our pion exchange kernel. This
needs to be further explored in future work.  
\begin{table*}[t]
 \begin{center}
 \small
\renewcommand{\arraystretch}{1.2}
  \begin{tabular}[h]{|c||c|c|c|}\hline
Nucleon &  RL1  &  RL2	        &	RL2 + $\pi$   \\ \hline\hline
 s-wave & 65.9  &  75.0 (1)		&	75.0 (1)     \\ \hline
 p-wave & 33.0  &  24.1 (3)		&	24.2 (0)     \\ \hline
 d-wave &  1.1  &   0.9 (1)		&	 0.8 (1)  \\ \hline
\end{tabular}\hspace{1cm}
  \begin{tabular}[h]{|c||c|c|c|}\hline
  Delta &  RL1 &  RL2	&	RL2 + $\pi$ \\ \hline\hline
 s-wave & 56.5 &  61.4 (15)	&	60.5 (14) \\ \hline
 p-wave & 39.9 &  31.0 (6)	&	31.1 (11) \\ \hline
 d-wave & 3.4  &  7.4 (20)	&	 8.1 (23) \\ \hline
 f-wave & 0.2  &  0.2 (1)	&	 0.3 (2) \\ \hline
\end{tabular}
\caption{Contribution in \% of the different partial wave sectors, at $m_{\pi}=138$~MeV, 
to the normalization of the Faddeev amplitudes for the Rainbow-Ladder kernel only (RL1) 
and for RL2 including pion cloud effects (RL2+$\pi$). As before, the numbers in brackets 
reflect the change of the results under variation of the interaction parameter $\eta$ 
between $1.6 \le \eta \le 2.0$. For RL1 this variation is very small and therefore no 
range is given. \label{tab:PartialWaveContributions}}
 \end{center}
\end{table*}

An observable effect of the slope of the mass-evolution curve close to the physical
point is given by the nucleon and delta sigma terms. In our approach, these are
trivially obtained using the Feynman-Hellman theorem
\begin{equation}
 \sigma_{\pi X}=m_q\frac{\partial M_X}{\partial m_q}\,,
\end{equation}\label{eq:sigma_terms}
where $m_q$ is the current-quark mass, $M_X$ is the baryon mass and the derivative 
is taken at the physical quark mass. For the nucleon we obtain 
\begin{eqnarray}
\sigma_{\pi N} &=& 30 (3)~\mbox{MeV (RL1)}, \nonumber\\
\sigma_{\pi N} &=& 26 (3)~\mbox{MeV (RL2)}, \nonumber\\
\sigma_{\pi N} &=& 31 (3)~\mbox{MeV (RL2+$\pi$)} \label{FH1}
\end{eqnarray}
for RL1, RL2 without and RL2 with pion exchange, respectively. Likewise, we obtain 
for the delta 
\begin{eqnarray}
\sigma_{\pi \Delta} &=& 24 (2)~\mbox{MeV (RL1)}, \nonumber\\
\sigma_{\pi \Delta} &=& 23 (3)~\mbox{MeV (RL2)}, \nonumber\\
\sigma_{\pi \Delta} &=& 24 (3)~\mbox{MeV (RL2+$\pi$)}\,. \label{FH2}
\end{eqnarray}
For the pion-nucleon case both of our values using physical parameters (RL1 and RL2+$\pi$) 
are slightly below the lower bound of a range of recent lattice 
results \cite{Shanahan:2012wh,Alvarez-Ruso:2013fza,Bali:2013dpa}. From a comparison of the
quark core calculation RL2 with RL2+$\pi$ we infer that about twenty percent of the nucleon
sigma term are generated by pion cloud effects. For the $\Delta$ this fraction is considerably 
smaller and our results in general are about 30 \% lower than available model 
results \cite{Lyubovitskij:2000sf,Cavalcante:2005mb}.

Within certain limits, the slope can be influenced by the choice of the model
parameters as reflected in the numbers in brackets given in (\ref{FH1}) and (\ref{FH2}).
However, as mentioned above, in order to study the mass evolution of the system 
and the resulting sigma-terms in more detail, one should include 
the effects of the gluon self-interaction in the two-body and three-body 
correlations, since these may have a significant impact \cite{Fischer:2009jm}. 
In addition, an improvement of the pion-exchange kernel by including terms 
with two-pion intermediate states 
(as discussed at the end of section \ref{subsec:pion_exchange} may have a
material impact on the slope close to the physical point and therefore result
in substantial changes in the sigma terms.

\subsection{Internal composition}\label{subsec:internal_composition}

Some insight into the internal structure of the baryon can be gained by 
studying the relative importance of the different partial-wave sectors. 
As shown in \cite{Eichmann:2009qa,Eichmann:2009en,SanchisAlepuz:2011jn}, 
Poincar\'e covariance enforces that in our framework baryons are composed, 
in principle, by s-, p- and d-wave components for spin-$\nicefrac{1}{2}$ 
particles and s-, p-, d- and f-wave components for spin-$\nicefrac{3}{2}$
particles. Therefore, one \textit{cannot} restrict the partial-wave 
composition in a covariant way and it is the dynamics what 
dictates the contribution of these components to a given state. Moreover, 
in the case of the nucleon, the flavor part of the Faddeev amplitude contains 
a mixed-symmetric and a mixed-antisymmetric term, as dictated by symmetry. 
Each of these is accompanied by a spin-momentum part; these are not identical
but related to each other. In our calculation we take all these contributions
into account. 

Form factors are observables which are expected to be more sensitive to the internal structure of the baryon. In particular, the $N\Delta\gamma$ transition \cite{Eichmann:2011aa,NDg} as well as the electromagnetic $\Delta$-baryon form factors \cite{Sanchis-Alepuz:2013iia} show a qualitatively different behavior when the angular-momentum content is artificially restricted. For this reason, we have calculated the contribution of the different partial-wave sectors to the normalization of the $N$ and $\Delta$ amplitudes when the pion corrections are or are not included, see Table \ref{tab:PartialWaveContributions}. In the case of the nucleon we average the contributions from the mixed-symmetric and mixed-antisymmetric terms. The angular-momentum composition of the state is not, nevertheless, the only element determining the form factors. The coupling of the photon (in case of electromagnetic form factors) and pion cloud plays an important role and is likely to be the dominant correction for, e.g., the baryon's charge radius and magnetic moment. This is, however, beyond the scope of this work.

Accepting the aforementioned caveats, it is nevertheless interesting to 
discuss the internal structure of the nucleon and $\Delta$ displayed in 
Tab.~\ref{tab:PartialWaveContributions}. Let us begin by analyzing 
the nucleon results. From comparison of our three setups it is clear 
that the inclusion of pion cloud effects induce only slight but potentially
significant changes in the angular-momentum content of the nucleon. These
are, however, not induced directly by the pion exchange term (cp. RL2 with
RL2+$\pi$), but by the accompanying change in the interaction scale of the
core rainbow-ladder contribution. In coordinate space this change of scale 
corresponds to a decrease of the core size, resulting in a larger s-wave 
component. This new balance is hardly affected by the explicit pion contributions.
It remains to be seen, how this affects the form factors of the nucleon.
Here, possible quantitative corrections will be dictated by the direct 
pion-photon interaction and may be large in the magnetic moments and the neutron
form factors at low momentum transfer \cite{Eichmann:2011vu}.
The case of the $\Delta$ is slightly different from the nucleon. Also here, the
main effects are generated by the modified interaction range of the core
rainbow-ladder contribution. The increase of the s-wave contributions as compared
to p-wave is less severe than in the nucleon case. Instead, the d-wave contributions
increase significantly with more than doubling their relative size as compared to
pure rainbow-ladder. This might have a significant impact in those form factors 
that measure the deformation of the $\Delta$-baryon, {\it i.e.} the electric quadrupole 
and the magnetic octupole \cite{Sanchis-Alepuz:2013iia}. Especially the latter 
one is small and therefore may be very sensitive to changes in the baryon internal 
structure.

\section{Summary}
\label{sec:summary}

In this work we included, for the first time, the explicit effects of pion cloud contributions 
in a description of baryons within a covariant three-body Faddeev approach. Previously,
these effects have been studied within a diquark-quark approximation of the Faddeev-equation
using an NJL-type interaction together with a perturbative treatment of the pion cloud effects 
\cite{Ishii:1998tw}.
In this study we have improved this calculation in three aspects: we solved the
genuine three-body equation, our interaction is much richer in terms of momentum dependence
and we treated the pion cloud effects non-perturbatively. Our approach generalizes the rainbow-ladder 
calculations of Refs.~\cite{Eichmann:2009qa,Eichmann:2009en,SanchisAlepuz:2011jn}
and complements corresponding efforts in the light meson sector 
\cite{Fischer:2007ze,Fischer:2008sp,Fischer:2008wy}. We found substantial 
contributions of the pion cloud effects to the masses of the baryons of 
the order of 5-15 \%, depending on the parameters of the underlying quark-gluon interaction.
In addition, we found slight but significant changes in the structure 
of the baryons reflected in the relative contributions of their partial waves.
We will explore the impact of these effects onto the electromagnetic 
as well as axial form factors of the baryons in future work.

\section*{Acknowledgments}
We are grateful to Gernot Eichmann and Richard Williams for discussions 
and a critical reading of the manuscript and to Walter Heupel for discussions. 
CF thanks the Yukawa Institute for Theoretical Physics, Kyoto University, 
where this work was completed during the YITP-T-13-05 workshop
on 'New Frontiers in QCD'. This work was supported by the Helmholtz 
International Center for FAIR within the LOEWE program of the State of 
Hesse, by the Helmholtz Center GSI, by the Erwin Schr\"odinger 
fellowship J3392-N20 of the FWF and by the DFG transregio TR 16.

\end{document}